\documentclass[aps, prb, reprint, superscriptaddress]{revtex4-2}

\usepackage[utf8]{inputenc}
\usepackage[T1]{fontenc}
\usepackage{newtxtext}
\usepackage{newtxmath}
\usepackage[scaled=0.92]{helvet}
\usepackage[english]{babel}
\usepackage{microtype}
\usepackage{xspace}
\usepackage{amsmath, amsfonts}
\usepackage{bm}
\renewcommand*{\epsilon}{\varepsilon}
\renewcommand*{\theta}{\vartheta}
\renewcommand*{\rho}{\varrho}
\renewcommand*{\phi}{\varphi}

\DeclareMathOperator\rp{Re}
\DeclareMathOperator\ip{Im}

\renewcommand{\vec}[1]{\bm{#1}}
\newcommand*{\hamil}{\mathcal{H}}
\DeclareMathOperator\tr{tr}
\newcommand*{\tensorial}[1]{\mathcal{#1}}
\usepackage[range-units=single, list-units=single, detect-all]{siunitx}
\DeclareSIUnit\rydberg{Ry}
\usepackage{graphicx}
\usepackage[autostyle]{csquotes}
\usepackage{booktabs}
\usepackage[breaklinks, hidelinks]{hyperref}

\providecommand*{\tab}{Tab.\@\xspace}
\providecommand*{\fig}{Fig.\@\xspace}
 \providecommand*{\ref}{Ref.\@\xspace}

\providecommand*{\eq}{Eq.\@\xspace}

\newcommand{\pll}{\phantom{\lambda_0}}
\newcommand{\twotofour}[2][]{	\makebox[0pt]{		\smash{\raisebox{.5\normalbaselineskip}{				#2}\hspace*{4\arraycolsep}}}}

\usepackage{pgf}
\usepackage{import}
\usepackage[version=4]{mhchem}

\graphicspath{{figures/}}
\newcommand*{\acronym}[1]{\textsc{\MakeLowercase{#1}}}
  \newcommand*{\iu}{\mathrm{i}} 
\renewcommand*{\star}{*}
\providecommand*{\DeltaJ}{\ensuremath\Delta J}

\newcommand*{\affilUKN}{  Fachbereich Physik,
  Universität Konstanz,
  D-78457 Konstanz,
  Germany}
\newcommand*{\affilWRC}{  Department of Theoretical Solid State Physics,
  Institute for Solid State Physics and Optics,
  Wigner Research Centre for Physics,
  H-1121 Budapest,
  Hungary}
\newcommand*{\affilBME}{  Department of Theoretical Physics, Institute of Physics,
  Budapest University of Technology and Economics,
  Műegyetem rkp.\ 3.,
  H-1111 Budapest,
  Hungary}
\newcommand*{\affilHUNREN}{  HUN-REN-BME Condensed Matter Research Group,
  Budapest University of Technology and Economics,
  Műegyetem rkp.\ 3.,
  H-1111 Budapest,
  Hungary}

\begin{document}

\title{Altermagnetic magnon splitting in hematite: from bulk g-wave to planar i-wave}

\author{Tobias Dannegger}
\email{tobias.dannegger@uni-konstanz.de}
\affiliation{\affilUKN}

\author{Julian A. Beisch}
\affiliation{\affilUKN}

\author{Levente Rózsa}
\affiliation{\affilWRC}
\affiliation{\affilBME}

\author{László Szunyogh}
\affiliation{\affilBME}
\affiliation{\affilHUNREN}

\author{Ulrich Nowak}
\affiliation{\affilUKN}

\date{\today}

\begin{abstract}
  Hematite is an insulating iron oxide with both a collinear antiferromagnetic and a weakly ferromagnetic phase with intriguing properties, such as long-range spin transport distances, that make it promising for spintronics applications.
  It has also been identified as a candidate for g-wave altermagnetism.
  Here, we present the results of fully relativistic ab initio calculations of the magnon properties of hematite, with a focus on altermagnetic effects.
  We quantitatively evaluate both the anisotropic and altermagnetic magnon band splitting as well as the magnon lifetimes.
  We show that dipolar interactions lead to a sizeable non-altermagnetic splitting and a hybridisation between opposite-chirality magnons resulting in a strong reduction of the spin expectation value of magnons close to  the $\Gamma$ point and around nodal planes, even in the collinear out-of-plane phase.
  Both the frequencies and the lifetimes have a large altermagnetic splitting in the $k_z = 0$ plane, which is characteristic of hematite and not generally expected for g-wave altermagnets.
  We compare our results to the prototypical, hexagonal g-wave altermagnet manganese telluride, which has four symmetry-protected nodal planes, whereas the fourth nodal surface in hematite is not completely fixed by symmetry.
  We derive implicit analytical expressions for this surface and discuss how it arises from a combination of rotational and translational symmetry, which leads to a transition to i-wave symmetry at the Brillouin zone boundary.
  The example of hematite reveals the trigonal altermagnets as a special case where the splitting function exhibits both bulk (g-wave) and planar (i-wave) components,
  and shows that the point symmetries alone are not sufficient to capture the full symmetry-determined behaviour of the splitting function.
\end{abstract}

\maketitle

\section{Introduction}

Hematite (\ce{\alpha-Fe2O3}) is an insulating iron oxide whose many intriguing magnetic properties have made it an important research subject in the field of magnetism.
The Morin transition from its collinear antiferromagnetic ground state to a canted, weakly ferromagnetic state at temperatures above $T_\text{M} = \qty{250}{\kelvin}$~\cite{Morin1950, MorrishBook} led to the theory of antisymmetric exchange, now known as the Dzyaloshinskii-Moriya interaction (\acronym{DMI})~\cite{Dzyaloshinskii1958, Moriya1960a}.
Its popularity in antiferromagnetic spintronics is due to its remarkably low magnetic damping, which allows for long-range spin transport~\cite{Lebrun2018}, and its unusally high Néel temperature of almost $\qty{1000}{\kelvin}$.
In 2022, it has been identified as a potential altermagnet~\cite{Smejkal2022_Hematite}.

Altermagnetism is a class of collinear magnetism differentiated from other antiferromagnets by its lack of certain symmetries.
Namely, a time-reversal operation combined with either a translation or inversion maps a non-altermagnetic lattice onto itself.
In reciprocal space, this symmetry ensures the degeneracy of opposite-spin bands at every $k$-point in the nonrelativistic limit.
Altermagnets are defined by the absence of such symmetries, which allows for a large, nonrelativistic spin splitting everywhere in $k$-space except on certain \emph{nodal surfaces}, where the splitting function (the energy difference between opposite-spin bands) changes sign.

These statements can be formalised using spin groups~\cite{LitvinOpechowski1974_SpinGroups,Litvin1977_SpinGroups}, which contain decoupled spin-space and real-space symmetry transformations, describing the symmetry of magnetic systems without spin-orbit coupling.
For collinear magnetic structures, all possible spin-Laue groups can be sorted into three types: ferro- and ferrimagnets (type I), non-altermagnetic antiferromagnets (type II), and altermagnets (type III)~\cite{Smejkal2022_Hematite}.
There is no special name such as \enquote{altermagnetic} for the type-II magnets, so in the following we will use the term \enquote{equimagnetic}, because opposite-spin bands in $k$-space are equal rather than alternating.

The type-III magnets (altermagnets) can be sub-categorised further, based on the spin-Laue group, into d-wave, g-wave, and i-wave altermagnets, depending on the number of symmetry-imposed nodal surfaces in the spin splitting (two, four, or six, respectively).
The configuration of the nodal surfaces additionally differentiates \emph{planar} altermagnets (if all nodal surfaces contain the primary symmetry axis) or \emph{bulk} altermagnets (if in-plane nodal surfaces exist)~\cite{Smejkal2022_Hematite}.

Hematite belongs to the bulk g-wave group.
But its symmetry with the spin-Laue group $^{1}\bar{3} ^{2}$m differs from that of other well-known altermagnets in this group, such as \ce{MnTe}~\cite{Krempasky2024, Liu2024_MnTe} and \ce{CrSb}~\cite{Zhang2025_CrSb}, which both have the nickeline (B8\textsubscript{1}) structure with spin-Laue group $\mathrm{^{2}6/^{2}m ^{2}m ^{1}m}$.
For hematite, the presence altermagnetic magnon band splitting has been predicted~\cite{Smejkal2023_ChiralMagnons, Hoyer2025_Hematite} and measured~\cite{Sun2025_Hematite}.

Here, we investigate the altermagnetic magnon band structure of hematite based on fully relativistic ab initio-calculated exchange interactions.
This allows us to compare altermagnetic and relativistic effects quantitatively.
The former is caused by the reduced symmetry of the isotropic part of the exchange interaction, which results in a sizeable magnon band splitting close to the Brillouin zone edge.
The non-isotropic terms in the Hamiltonian lead to a much smaller band splitting but they have a strong effect on the magnons close to the Brillouin zone centre.

\section{Spin model}

Our calculations are based on the Heisenberg spin model for hematite developed in Ref.~\cite{HematiteAbInitio}.
This model is built on density functional theory (\acronym{DFT}) calculations using the Perdew-Burke-Ernzerhof functional within a generalised gradient approximation with Hubbard correction.
The electronic band structure was computed with the screened Korringa--Kohn--Rostoker (\acronym{SKKR}) multiple scattering method and mapped onto bilinear exchange tensors by means of spin-cluster expansion.
Together with the two uniaxial anisotropy parameters $d_{2}$ and $d_{4}$, derived by fitting the model to experimental spin-flop measurements, it was shown to reproduce the equilibrium properties of hematite with good quantitative accuracy.
The spin Hamiltonian is given by
\begin{equation}
  \hamil = -\tfrac{1}{2} \sum_{i \neq j} \vec{S}_{i} \tensorial{J}_{ij} \vec{S}_{j} - \sum_{i} \left[ d_{2} S_{i,z}^{2} + d_{4} S_{i,z}^4 + \mu \vec{S}_{i} \cdot \vec{B} \right]
  ,
  \label{eq:Hamiltonian}
\end{equation}
with exchange tensors $\tensorial{J}_{ij}$, anisotropy energies of second and fourth order, $d_{2}$ and $d_{4}$, the external magnetic field~$\vec{B}$, and the atomic magnetic moments with magnitude $\mu \approx \num{4.2} \mu_{\mathrm{B}}$ represented by unit vectors $\vec{S}_{i}$.

We then use linear spin-wave theory to analyse the magnon dispersion.
We employ both a linearised Landau--Lifshitz--Gilbert (\acronym{LLG}) equation as well as a linearised Holstein--Primakoff transformation.
In the damping-free linear limit, both methods are in agreement.
We mainly use the former in order to include the effect of Gilbert damping, with a rather low damping constant of $\alpha = \num{e-5}$, but turn the latter for calculations of the spin expectation value and to derive analytical expressions for the nodal surfaces.

\subsection{Altermagnetic structure}

\begin{figure}
  \centering
  \includegraphics{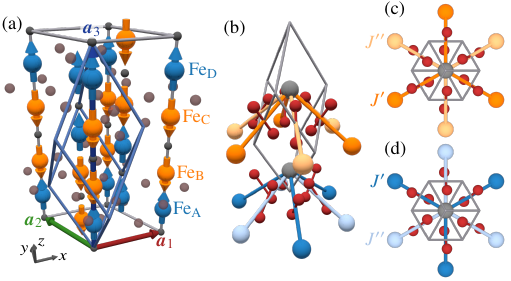}
  \caption{    (a) Crystal structure of hematite in the conventional hexagonal (grey) and primitive rhombohedral (blue) unit cells.
    The blue and orange spheres are the \ce{Fe^3+} ions with magnetic moments aligned parallel or antiparallel to the $c$-axis.
    The smaller spheres symbolise nonmagnetic oxygen atoms.
    (b--d) Visualisation of the 13\textsuperscript{th}-nearest neighbours of an iron atom (grey sphere in the centre) with exchange energies $J'$ (darker spheres) and $J''$ (lighter spheres).
  }
  \label{fig:UnitCellShellPairs}
\end{figure}

Hematite crystallises in the trigonal corundum (D5\textsubscript{1}) structure with space group {R$\bar{3}$c} (no.~167), as shown in \fig~\ref{fig:UnitCellShellPairs}(a).
The primitive rhombohedral unit cell contains four magnetic \ce{Fe^3+} sites and six nonmagnetic \ce{O^2-} sites.
Due to the presence of the oxygen atoms, the 13\textsuperscript{th}-nearest Fe-Fe coordination shell by distance decomposes into two inequivalent sub-shells, as shown in \fig~\ref{fig:UnitCellShellPairs}(b), that are not related by any symmetry operation.
Therefore, the isotropic exchange energies $J'$ and $J''$ in these sub-shells are different because the arrangement of the nonmagnetic atoms facilitating the interaction differs along their connecting lines.

Crucial for altermagnetism is the way that these neighbours are arranged on the opposite-spin sublattices.
For example, the neighbourhood of sublattice~A, shown in \fig~\ref{fig:UnitCellShellPairs}(d), is related to that of sublattice~C, see \fig~\ref{fig:UnitCellShellPairs}(c), by a glide plane symmetry.
Similarly, sublattice~A can be mapped onto sublattice~B by a \qty{180}{\degree} rotation.
The parallel sublattices A and D, by contrast, are connected by an inversion symmetry.
This differentiates hematite from \ce{Cr2O3}, which has the same structure but with a G- instead of A-type magnetic structure, making it non-altermagnetic~\cite{Corliss1965_Cr2O3}.

\subsection{Exchange graph}

Since altermagnetism is a spin-group property, it can only arise from those terms in the spin Hamiltonian that depend on the relative orientation of spins but not on the absolute orientation relative to the lattice.
In \eq~\eqref{eq:Hamiltonian}, those are the isotropic parts of the exchange interactions, $J_{ij} = \tfrac{1}{3}\tr \tensorial{J}_{ij}$.
The ab initio calculations were performed up to the 34\textsuperscript{th}-nearest neighbours (equivalent to two hexagonal in-plane lattice constants).
This amounts to a total of 170 neighbours per iron atom, which are fully described by 42 unique exchange couplings (the rest can be derived by symmetry operations).
Of the 42 isotropic exchange energies, twelve exhibit the altermagnetic symmetry breaking described above.
These are printed in \tab~\ref{tab:AltermagneticExchangeEnergies}.
\begin{table}
  \caption{List of all exchange interactions up to the 34\textsuperscript{th}-nearest neighbours that exhibit altermagnetic symmetry breaking.
  Printed are the ordinal number of the Fe-Fe coordination shells in the first column, then the number of neighbours (n.n.) in that shell, the sublattices to which the interaction partners belong (relative to sublattice~A), and then the two different exchange integrals $J'$ and $J''$ computed by spin-cluster expansion for these neighbours, as well as their difference $\Delta J$.
  }
  
\begin{tabular}{%
  S
  S
  l
  S
  S
  S
  }
\toprule
    {\hspace*{-1em}no.}
  & {n.n.}
  & {subl.}
  & {$J'$ (\unit{\micro\electronvolt})}
  & {$J''$ (\unit{\micro\electronvolt})}
  & {$\Delta J$ (\unit{\micro\electronvolt})} \\
\midrule
13 &  6 & A--D & -2684.1 &   118.5 &  2802.6 \\
17 &  6 & A--D &    -9.8 &   -11.5 &     1.7 \\
20 & 12 & A--C &   -21.2 &    -1.2 &    20.0 \\
22 & 12 & A--C &   -23.6 &    -1.5 &    22.1 \\
25 & 12 & A--A &    -2.2 &     0.5 &     2.7 \\
28 &  6 & A--D &    -2.6 &    -3.3 &     0.6 \\
\bottomrule
\end{tabular}

  \label{tab:AltermagneticExchangeEnergies}
\end{table}
The full list of all exchange interactions in the spin Hamiltonian is available in Ref~\cite{HematiteSpinModelData}.

By far the largest altermagnetic interactions, both in terms of absolute magnitude and splitting, are those in the 13\textsuperscript{th} shell, which are also the first ones.
In fact, $J_{13}'$ is not only the largest altermagnetic interaction but also the largest interaction at all beyond the fourth-nearest neighbours.
For the 13\textsuperscript{th} shell, our ab initio calculations suggest a remarkably large splitting of $\Delta J = \qty{2.8}{\milli\electronvolt}$.

This is similar to the values found for the g-wave altermagnets MnTe~\cite{Liu2024_MnTe} and CrSb~\cite{Zhang2025_CrSb}, with which we compare our results in \tab~\ref{tab:ExchangeComparison}.
\begin{table}
  \caption{Comparison of the altermagnetic exchange interaction splitting of hematite with that of MnTe~\cite{Liu2024_MnTe} and CrSb~\cite{Zhang2025_CrSb}.
  The number of the first altermagnetic coordination shell is printed in the second column, followed by the number of neighbours (n.n.) in that coordination shell, the neighbour distance~$d$ of that shell, and the different exchange integrals $J'$ and $J''$, as well as their difference $\Delta J$.
  }
  
\begin{tabular}{%
  l
  r
  r
  S
  S
  S
  S
  }
\toprule
    {mat.}
  & {\hspace*{-1em}no.}
  & {n.n.}
  & {$d$ (\unit{\angstrom})}
  & {$J'$ (\unit{\milli\electronvolt})}
  & {$J''$ (\unit{\milli\electronvolt})}
  & {$\Delta J$ (\unit{\milli\electronvolt})} \\
\midrule
  \ce{Fe2O3} & 13 &  6 & 6.466 & -2.68 &  0.12 &  2.80 \\
  \ce{MnTe}  & 10 & 12 & 9.916 & -0.60 &  0.19 &  0.79 \\
  \ce{CrSb}  & 11 & 12 & 8.989 &  0.42 &  1.10 &  0.68 \\
\bottomrule
\end{tabular}

  \label{tab:ExchangeComparison}
\end{table}
With the coordination number taken into account, the total altermagnetic exchange splitting in hematite is \qty{77}{\percent} larger in hematite than in MnTe.
In MnTe and CrSb, the altermagnetic symmetry breaking first shows up in the 10\textsuperscript{th}- and 11\textsuperscript{th}-nearest interaction shells, respectively.
These are, however, much further away (approximately \qty{9}{\angstrom}) than the 13\textsuperscript{th}-nearest neighbours in hematite (\qty{6.5}{\angstrom}), which makes it somewhat plausible that this interaction energy would be larger than that of the other two.
In hematite, even the 30\textsuperscript{th}-nearest neighbours are only \qty{9.65}{\angstrom} away.
Nevertheless, our \acronym{DFT} calculations suggest that all altermagnetic interactions beyond the first one are at most of the order of \qty{20}{\micro\electronvolt} and may therefore be neglected.

By fitting to inelastic neutron scattering measurements, Sun et al.~\cite{Sun2025_Hematite} found an effective splitting of approximately \qty{0.97}{\milli\electronvolt}.
These experiments were performed at \qty{50}{\kelvin}, so thermal softening of the magnon modes may account for some of the reduction compared to our zero-temperature calculations.

\section{Magnon dispersion relation}

\subsection{Altermagnetic frequency splitting}

\begin{figure}
  \centering
  \includegraphics{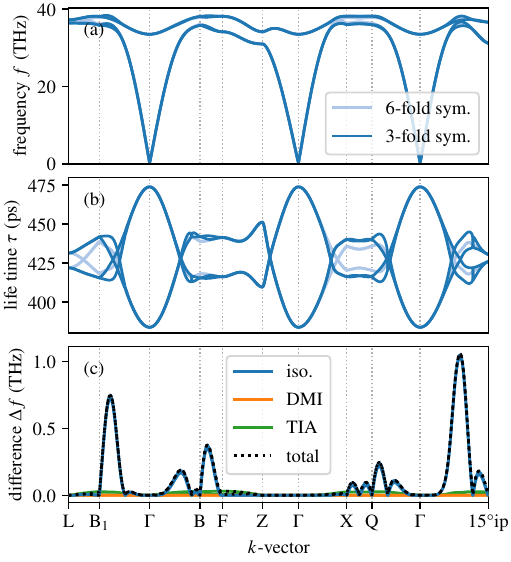}
  \caption{    Magnetic dispersion relation for hematite along the high-symmetry path compared for a spin model with the three-fold rotational symmetry and an artificial six-fold rotational symmetry in the 13\textsuperscript{th} interaction shell.
    Plotted are the magnon frequency (a) and lifetime (b), as well as the frequency splitting (c).
    The coloured lines in panel (c) show the contributions of different parts of the tensorial exchange to the total splitting, which is shown as the dashed black line.
    For the positions of the labelled high-symmetry points, cf.\ \fig~\ref{fig:NodalSurface}(a).
  }
  \label{fig:Dispersion}
\end{figure}

The calculated magnon dispersion relation for hematite is shown in \fig~\ref{fig:Dispersion}(a) for a specific path through the Brillouin zone.
The corresponding magnon lifetimes are plotted in \fig~\ref{fig:Dispersion}(b).

Because the unit cell contains four magnetic atoms, the spin-wave dispersion consists of four branches, grouped into two nearly degenerate pairs.
The splitting within each pair of branches in shown in \fig~\ref{fig:Dispersion}(c).
The altermagnetic splitting is the same in both pairs but the non-isotropic splitting differs between them and is usually larger for the lower-frequency branches.

The complete splitting vanishes exactly at the high-symmetry L~point and on the $\Lambda$ line ($\Gamma$Z).
Everywhere else the degeneracy is lifted by a combination of the \acronym{DMI} and the two-ion anisotropy (\acronym{TIA}), as well as the altermagnetic asymmetry of the isotropic exchange energies.

To determine how much of the splitting is caused by each of these three contributions, we selectively remove the individual terms from the Hamiltonian and compare the resulting amount of splitting to that of the full Hamiltonian.
To remove the altermagnetic splitting from the model, we symmetrise the exchange graph with an additional six-fold rotational symmetry that connects all six interactions in the 13\textsuperscript{th} coordination shell and replaces them with the mean of their originally different exchange energies.
We thereby turn the spin model into that of an accidental type-II magnet, whose band degeneracies are not enforced by symmetry.
The dispersion of this model is shown as the lighter blue lines in \fig~\ref{fig:Dispersion}(a--b).

To remove the \acronym{DMI}, we subtract the antisymmetric components from the individual exchange tensors.
Applying both symmetrisations leaves only the \acronym{TIA}-induced splitting.
The on-site anisotropies do not cause any splitting in the low-temperature phase because the magnetic moments in this collinear ground state are parallel to the easy axis.

The largest amount of splitting in the entire Brillouin zone reaches just above \qty{1}{\tera\hertz}, compared to magnon frequencies that reach roughly \qty{40}{\tera\hertz}.
In relative terms, this corresponds to a maximum splitting of \qty{2.9}{\percent}.
The analysis in \fig~\ref{fig:Dispersion}(c) shows that this splitting originates almost exclusively from the altermagnetic splitting of the isotropic exchange energies.
The non-isotropic splitting is only larger than the altermagnetic splitting on path segments such as $\Gamma$-Z-F, which coincides with an altermagnetic nodal plane, as we will discuss below.
Interestingly, the splitting reaches its maximum for $k$-vectors at \qty{15}{\degree} to the in-plane crystal axes, which does not correspond to any lattice plane with rational Miller indices.
In rhombohedral coordinates, this direction corresponds to $(-2-\sqrt{3}, 1+\sqrt{3}, 1) \approx (-26, 19, 7)$.
We will discuss the peculiar symmetry of the splitting function further below.

\subsection{Non-isotropic effects}

Of the non-isotropic terms, the \acronym{TIA} is the most relevant contribution.
For a correct description of the non-isotropic band splitting, these terms, which stem mostly from dipole-dipole interaction, must therefore be taken into account, while the \acronym{DMI} terms have very little effect in the low-temperature collinear phase.
Comparing the contributions of the three parts of the exchange tensors, we can say that the altermagnetic splitting reaches the \unit{\tera\hertz} range, the \acronym{TIA} causes an additional splitting in the \unit{\giga\hertz} range, and the \acronym{DMI}-induced splitting, while still non-zero, reaches only the \unit{\kilo\hertz} range.
The maximum non-isotropic splitting is found at the F~point and reaches \qty{31.06}{\giga\hertz}.

Given the sizeable non-isotropic splitting caused by the \acronym{TIA}, we take a closer look at its effect on the magnon chirality.
In previous works~\cite{Hoyer2025_Hematite, Hu2026_Hematite}, it was assumed that the $\pm \hbar$ spin expectation value of the magnons would always be preserved in the easy-axis phase, since the spins are aligned along the on-site anisotropy axis and two-site terms were not considered.
Our fully relativistic model, however, shows that the suppression of the magnon spin close to the $\Gamma$ point, which was predicted by Hoyer et al.~\cite{Hoyer2025_Hematite} for the easy-plane phase, is to be expected in the easy-axis phase as well, where it is caused by the \acronym{TIA}.
\begin{figure}
  \centering
  \includegraphics{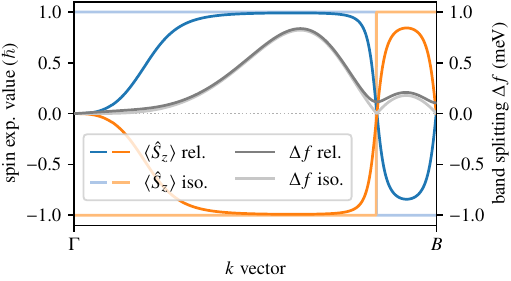}
  \caption{    The spin expectation value calculated for the path $\Gamma \to B$ for the acoustic magnon modes,
    once with all the interactions and once in the isotropic limit, with only on-site anisotropies and the exchange tensor included.
    Additionally the splitting of the acoustic magnon bands along the same path is shown.
  }
  \label{fig:SpinExpValue}
\end{figure}

 The spin expectation value $\langle \hat{S}_z \rangle_m(\vec{k})$ of each magnon mode $m$ is obtained from
\begin{align}
	\langle \hat{S}_z \rangle_m(\vec{k})
	&= \Gamma_m^\dagger(\vec{k}) S_z \Gamma_m(\vec{k}),\\
	S_z &= \mathrm{diag}\left(-1,\phantom{-}1,\phantom{-}1,-1,-1,\phantom{-}1,\phantom{-}1,-1\right).
\end{align}
Here, $\Gamma_m(\vec{k})$ denotes the $m$-th column of the Bogoliubov transformation matrix $\Gamma(\vec{k})$, which diagonalises the bosonic spin-wave Hamiltonian using Colpa's method~\cite{Colpa1978}.
In \fig~\ref{fig:SpinExpValue}, the spin expectation value along the quantisation axis, $\langle \hat{S}_z\rangle$, is plotted for the two lower-frequency (acoustic) magnon modes along the path $\Gamma$B, both with the non-isotropic interactions included and in the isotropic limit.
The non-isotropic two-site interactions show a strong effect on the nature of the magnon modes close to the $\Gamma$ point, but also at another point close to B, where the bands cross in the isotropic limit.
With non-isotropic interactions, this crossing turns into an anticrossing with hybridisation of the opposite-spin modes.
In agreement with previous work~\cite{Hoyer2025_Hematite}, we find that the quantised magnon spin is mostly recovered around lattice momenta where significant altermagnetic splitting occurs.

Although the term \enquote{relativistic} is often used for the entire non-isotropic part of the Hamiltonian, we want to highlight that the largest part of the non-altermagnetic splitting in hematite is not due to spin-orbit coupling but due to dipole-dipole interactions.
The purely relativistic (spin-orbit-mediated) \acronym{TIA} only accounts for a splitting of \qty{1.3}{\giga\hertz}, while the remaining \qty{30}{\giga\hertz} are of dipolar origin.
We therefore use the term \enquote{non-isotropic} here to emphasise that the largest non-altermagnetic splitting is also of non-relativistic origin.

\subsection{Lifetime splitting}
The magnon lifetimes, shown in \fig~\ref{fig:Dispersion}(b), become important when considering spin transport effects, as they determine the transport range for each mode.
Since hematite is known to have a very low Gilbert damping~\cite{Lebrun2018} (as mentioned above, we use $\alpha = \num{e-5}$), the magnon lifetimes are consequently very large, around half a nanosecond.
Unlike in ferromagnets, where the lifetimes are expected to be roughly inversely proportional to the frequencies, $\tau \propto (\alpha \omega)^{-1}$, the relationship in an antiferromagnet like hematite is more complicated and in this case, they are comparatively constant (varying by roughly $\pm\qty{10}{\percent}$).
The splitting between the lifetimes of adjacent branches reaches a maximum of approximately \qty{10}{\pico\second}.

\section{Symmetries of the splitting function}
\subsection{Comparison to other bulk g-wave altermagnets}

Having established the key magnon properties and their quantitative behaviour, including relativistic and nonrelativistic splitting,
we now discuss the symmetry of the altermagnetic splitting function $\Delta f_{i}(\vec{k}) = f_{i}^{-}(\vec{k}) - f_{i}^{+}(\vec{k})$, where $f_{i}^{-}(\vec{k})$ denotes the frequencies of the magnons with spin in negative $z$-direction at wave-vector $\vec{k}$ and $f_{i}^{+}(\vec{k})$ denotes the branches with spin in positive $z$-direction.
The index~$i$ distinguishes between the low-frequency ($i=1$) and high-frequency ($i=2$) branches.
We only consider isotropic terms here, so the magnon spin remains a good quantum number and we have $\Delta f_{1}(\vec{k}) = \Delta f_{2}(\vec{k}) =: \Delta f(\vec{k})$.

To observe the altermagnetic splitting, it is necessary to look at the regions of reduced symmetry.
Along the standard high-symmetry path through the Brillouin zone, given in Ref.~\cite{SetyawanCurtarolo2010} as $\Gamma$–L–B$_1$|B–Z–$\Gamma$–X|Q–F–P$_1$–Z|L–P, only non-altermagnetic splitting can be observed.
Part of this high-symmetry path are for example the sections $\Gamma$X and LB$_1$ plotted in \fig~\ref{fig:Dispersion}.
On the axis of highest symmetry in the crystal ($\Lambda = \Gamma$Z), the magnon branches become exactly degenerate with neither non-isotropic nor isotropic splitting.
Lower-symmetry lines such as $\Gamma$B and $\Gamma$B$_1$, by contrast, show a very large splitting of altermagnetic origin.

\begin{figure}
  \centering
  \includegraphics{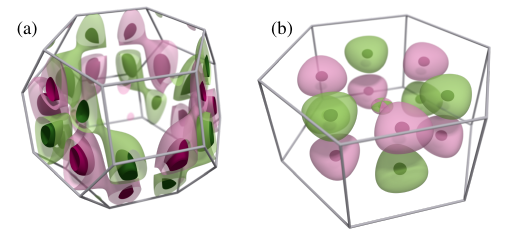}
  \caption{    Areas of largest splitting in the first Brillouin zone of (a)~hematite and (b)~MnTe.
    The shown contour surfaces indicate a splitting of \qty{400}{\giga\hertz} (light semi-transparent surfaces) and \qty{800}{\giga\hertz} (dark opaque surfaces).
    The $k_z = 0$ plane is indicated in grey.
  }
  \label{fig:Vol}
\end{figure}

A full three-dimensional representation of the areas of largest altermagnetic splitting within the rhombohedral Brillouin zone of hematite is shown in \fig~\ref{fig:Vol}(a).
Its classification as bulk g-wave suggests a six-lobe pattern with the chirality-split regions of the Brillouin zone separated by three nodal planes parallel to the rotation axis and a fourth nodal plane perpendicular to it.
This pattern can for examples be observed in manganese telluride, for which the splitting function, based on the spin model by Liu et al.~\cite{Liu2024_MnTe}, is plotted in \fig~\ref{fig:Vol}(b).

\begin{figure}[t]
  \centering\vspace*{1em}
  \hspace*{-6pt}\includegraphics{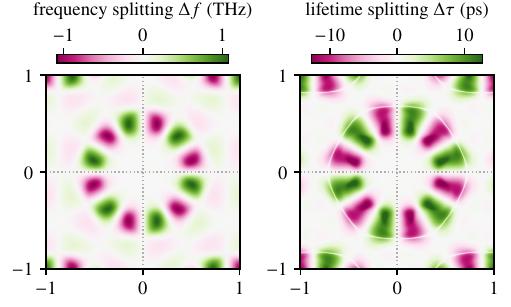}
  \caption{Altermagnetic splitting of the (a) frequencies and (b) lifetimes of the magnon bands in hematite in a cross-section of $k$-space with $k_z = 0$ computed from linear spin-wave theory with a Gilbert damping constant of $\alpha = \num{e-5}$.
  }
  \label{fig:DispersionZmap}
\end{figure}

For hematite, however, the splitting function is somewhat less symmetric.
It has the same set of three vertical nodal planes but the fourth nodal surface has a nontrivial shape determined by the exact model parameters.
And while the band splitting in MnTe vanishes on the boundaries of the Brillouin zone, the largest splitting in hematite can actually be observed directly at the zone edges.
The main difference, however, is that hematite has a finite in-plane splitting.
In fact, the splitting at $k_z = 0$ is the largest in the Brillouin zone.
The frequency and lifetime splitting inside this plane are plotted in \fig~\ref{fig:DispersionZmap}.
The points of largest splitting are observed within this cross-section at \qty{15}{\degree} angles to the hexagonal in-plane crystal axes $(1000) = (1\bar{1}0)$ in rhombohedral coordinates, which is also the last section of the path plotted in \fig~\ref{fig:Dispersion}.

Like in the hexagonal altermagnets, the splitting function in hematite still posesses a fourth nodal surface, which is, however, not planar but alternates between points above and below the basal plane, resulting in a twelve-fold pattern of alternating splitting inside the $k_z = 0$ cross-section, which is plotted in \fig~\ref{fig:DispersionZmap}.
Half of the nodes of this pattern come from the three trivial nodal planes that arise from the glide plane symmetries.
The other half of the nodes stem from the two-fold rotational symmetry axes, which also map opposite-spin sublattices onto each other and are perpendicular to the glide planes.
From a continuity argument, those axes must be connected by a fourth nodal plane~\cite{Urru2025_BFO}, but since only the two-fold axes themselves are symmetry-protected, the surface in-between the axes can take a variable shape, depending only on the exchange interactions.
In the following, we want to derive an implicit analytical expression for this surface.

We leave the details of the derivation to App.~\ref{appendix:HP}, but it makes use of the characteristic polynomial \eqref{eq:charpoly} of the eigenproblem of the linearised Holstein--Primakoff transformation.
This polynomial is a depressed quartic, so its eigenvalues are non-degenerate only if the coefficient of the linear term $c_3 \neq 0$.
By searching for the roots of $c_3$ in the Brillouin zone, one finds all nodal planes.

The $c_3$ coefficient can be expressed with five parameters $a, b, c, d, e \in \mathbb{C}$, which follow from the Hamiltonian parameters as detailed in App.~\ref{appendix:HP}, as
\begin{equation}
  c_3 = \frac{2}{\iu \hbar}  \left( b^\star  \left(c^\star  d   +  e^\star c \right)- \text{c.c.} - a \left( e^\star e  - d^\star d\right)     \right)
  .
\end{equation}
In the case of the artifically equimagnetic model, one can calculate that $d^\star = e$ for every $k$-point.
This is already a root of $c_3$ and causes the magnon band degeneracy across the whole spectrum.

\begin{figure}
  \centering
  \includegraphics{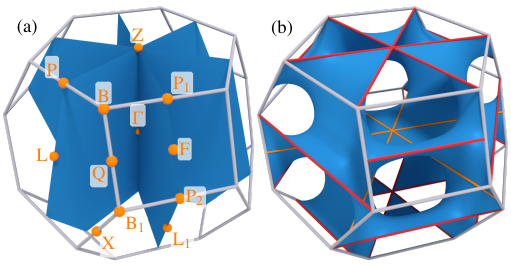}
  \caption{    (a) The three glide-plane-protected nodal planes in hematite shown inside the first Brillouin zone.
    The conventional high-symmetry points are labelled in orange.
    (b) The fourth nodal surface, protected by twofold rotational symmetries.
    The orange lines are the three twofold rotation axes going through the $\Gamma$ point.
    The red lines are additional twofold axes obtained through the inclusion of translational symmetry, which pass through the boundary facets of the first Brillouin zone.
  }
  \label{fig:NodalSurface}
\end{figure}

For the altermagnetic case, with only a three-fold rotational symmetry, the roots can be found analytically by a reformulation of the $c_3$ equation:
\begin{widetext}
\begin{gather}
    c_3 \propto \DeltaJ \cdot \left[ \sin\left(  {k}^{x}\vec{\Delta}^{1,x}_{AD} +{k}^{y}\vec{\Delta}^{1,y}_{AD}\right)  +\sin\left(  {k}^{x}\vec{\Delta}^{2,x}_{AD} + {k}^{y}\vec{\Delta}^{2,y}_{AD}\right) +\sin\left( {k}^{x}\vec{\Delta}^{3,x}_{AD} + {k}^{y}\vec{\Delta}^{3,y}_{AD}\right) \right] \cdot \\
    \cdot \Bigg[ \sum_l J^l_{AC} \cos \left(  \vec{k} \vec{\Delta}^l_{AC}\right)
    \cdot \sum_l J^l_{AB} \sin \left(  \vec{k} \vec{\Delta}^l_{AB}  - {k}^z\vec{\Delta}^{z}_{AD}\right)
       - \left( \sum_l J^l_{AA} \cos \left(  \vec{k} \vec{\Delta}^l_{AA}\right) +F \right)
    \cdot   \sum_{l\leq 10} J^l_{AD} \sin \left(  \vec{k} \vec{\Delta}^l_{AD} - {k}^z\vec{\Delta}^{z}_{AD}\right) \Bigg],
\end{gather}
\end{widetext}
with
\begin{equation}
F = - \sum_l J^l_{AA} + \sum_l J^l_{BA} +  \sum_l J^l_{CA} - \sum_l J^l_{DA}.
\end{equation}
The roots of the first bracket, containing three sine terms, give the three trivial nodal planes.
These planes are independent of the $z$-componenent of the wave vector and can be seen in \fig~\ref{fig:NodalSurface}(a).

The roots of the second bracket generate the fourth nodal plane.
Its precise shape is parametrised by the pair distance vectors $\vec{\Delta}$ and their exchange energies $J$.
For our ab initio parameters, the result is shown in \fig~\ref{fig:NodalSurface}(b).
The actual shape can vary quite drastically for different exchange energies (an interactive visualisation tool is available at~\cite{NodalSurfaceVisualiser} to try out different parameters), but it always goes through the two-fold rotational symmetry axes that guarantee its existence.

The three two-fold axes contained in the Laue group $^1\bar{3}^2$m, shown in orange in \fig~\ref{fig:NodalSurface}(b), lie in the $k_z = 0$ plane and cross each other at the $\Gamma$ point, which causes the nodal surface to approximate the basal plane there as well.
To understand the behaviour of this nodal surface away from the $\Gamma$ point, however, we must go beyond the point symmetries and consider the periodicity of the reciprocal lattice as well.
Combining the two-fold rotational symmetries with reciprocal lattice point translations leads to additional rotation axes shown in red in \fig~\ref{fig:NodalSurface}(b).
These additional symmetry lines are not at $k_z = 0$ and cause the nodal surface to bend away from the basal plane, in $+k_z$-direction around the L points and in $-k_z$-direction around the L$_1$ points.

This alternating pattern, enforced by combined rotational and translational symmetries, warps the nodal surface into an almost vertical orientation close to the X points.
Together with the three mirror-protected nodal planes, these three additional vertical sections lead to the twelve-fold pattern observed in \fig~\ref{fig:DispersionZmap}.
In effect, the four nodal planes seen at $\Gamma$ look like six nodal planes near the zone edge, where the splitting is also largest.
Six nodal planes, however, are characteristic of i-wave altermagnets.
In case of only vertical planes, it would indicate planar i-wave altermagnetism~\cite{Smejkal2022_Hematite}.
This explains why the splitting pattern in \fig~\ref{fig:DispersionZmap} looks like that of a planar i-wave rather than a bulk g-wave altermagnet.
While the bulk g-wave pattern expected from the spin-Laue group analysis is recovered in the vicinity of the $\Gamma$~point, the splitting is actually largest at the points furthest away from the $\Gamma$~point, where the symmetry can only be understood when translational symmetries are taken into account.

\subsection{Spherical harmonics expansion}

To quantify this transition from the bulk g-wave to the planar i-wave pattern, we expanded the splitting function in terms of simple spherical harmonics on spheres of varying radius.
\begin{figure}
  \centering
  \includegraphics{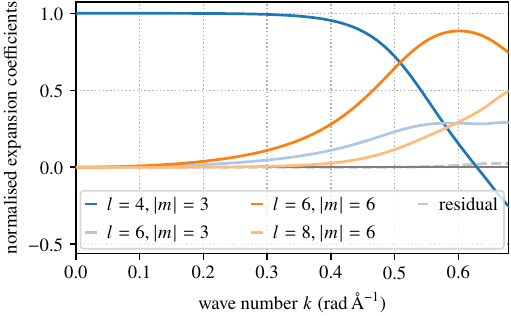}
  \caption{    Expansion of the splitting function in spherical harmonics, with expansion coefficients normalised to the function's $L_2$ norm.
    The dark blue and orange lines correspond to the leading bulk g-wave and planar i-wave components, respectively.
  }
  \label{fig:SHE}
\end{figure}
The result is shown in \fig~\ref{fig:SHE} for the four lowest-order allowed harmonics, with components normalised to the total amount of splitting.
At zero, this analysis shows a pure bulk g-wave pattern, which gains more and more planar i-wave character at higher momenta.
At $k = \qty{0.51}{\radian\per\angstrom}$, the i-wave component becomes larger than the g-wave and above \qty{0.6}{\radian\per\angstrom}, it is by far the dominant component.

This result does not negate the categorisation of hematite as a bulk g-wave altermagnet, which is based purely on point symmetries and therefore describes the behaviour at low wave numbers, but it highlights the need to go beyond the textbook picture of the g-wave pattern when dealing with the trigonal members of this class.
This is particularly important in this case because the spin-Laue group $^1\bar{3}^2$m is the only one that allows for both bulk and planar components simultaneously.
As we have seen with hematite, this can mean that the supposedly nodal basal plane is in fact the one where the largest band splitting can be found, rather than along the out-of-plane directions one would investigate first in materials like MnTe and CrSb.

\section{Conclusions}

To conclude, we have presented ab initio calculations that show an altermagnetic splitting of \qty{2.8}{\milli\electronvolt} in the isotropic exchange interactions of the 13\textsuperscript{th} Fe-Fe coordination shell in hematite, substantially larger than that of other g-wave altermagnets, resulting in a magnon band splitting above \qty{4}{\milli\electronvolt} (which is similar to but slightly larger than previous estimates of \qty{2}{\milli\electronvolt}~\cite{Hoyer2025_Hematite}, \qty{3}{\milli\electronvolt}~\cite{Smejkal2023_ChiralMagnons}, and \qty{3.67}{\milli\electronvolt}~\cite{Hu2026_Hematite}).

Our fully relativistic spin model showed that dipole-dipole interactions lead to the largest non-altermagnetic magnon band splitting and to significant hybridisation of opposite-spin branches close to the Brillouin zone centre, even in the easy-axis phase, while spin-orbit coupling only leads to negligible effects in comparison.

The altermagnetic splitting of the magnon frequencies and lifetimes exhibits a nontrivial pattern that results in a planar-i-wave-like arrangement of {twelve} antinodes in the $k_z = 0$ plane, which stands in strong contrast to other g-wave altermagnets such as \ce{MnTe}, where $k_z = 0$ coincides with a nodal plane protected by mirror symmetry.
The additional symmetry breaking in hematite would allow for in-plane spin transport, which is forbidden by the symmetry of MnTe and CrSb.

We derived an analytical expression for the nontrivial fourth nodal surface in hematite and showed that its complex shape can only be understood when translational symmetries are taken into account.
These enforce an azimuthally alternating chirality, which, together with the three trivial nodal planes, fixes the i-wave pattern on the zone boundary.
Because the splitting is strongest at the very edge of the Brillouin zone, the i-wave pattern is the dominant contribution to the splitting function in hematite.

These results show the need to go beyond the paradigm of point symmetries in the classification of altermagnets and consider the translational symmetries of the reciprocal lattice.
Instead of only asking which component is allowed at $k \to 0$, where the altermagnetic splitting vanishes, we should assess which components are allowed by symmetry at all.
The example of hematite highlights that the lowest-order component is not necessarily the largest and the Brillouin zone's boundary conditions may indeed enforce a higher harmonic than would be expected from an analysis of the Laue group.

\smallskip
\begin{acknowledgments}
  This work has been supported by the German Research Foundation (DFG) under project No.~581436551.
  LR and LS acknowledge additional support by the National Research, Development and Innovation Office under No.~K142652, FK142601 and \acronym{ADVANCED}~149745.
\end{acknowledgments}

\section*{Data availability}

The data that support the findings of this article are openly available~\cite{HematiteSpinModelData, AltermagneticHematiteDispersionData}.

\appendix

\section{Calculation details}\label{appendix:HP}

We used two different methods to obtain the magnon dispersion relation: through a linearised Holstein--Primakoff transformation, similar to Ref~\cite{Hoyer2025_Hematite}, and via the linearised Landau--Lifshitz--Gilbert (\acronym{LLG}) equation of motion.
The \acronym{LLG} approach starts out directly from the semiclassical description and has the advantage that the obtained eigenvalues $\lambda$ contain not only the magnon frequency $\omega = \ip(\lambda)$, but also their lifetime $\tau = -\rp(\lambda)^{-1}$.
The Holstein--Primakoff approach starts from a quantum mechanical spin model and takes the classical limit by neglecting higher-order terms.
The quantum-mechanical version of the spin model ($q$) requires a rescaling of the classical parameters ($c$) in the Hamiltionian \eqref{eq:Hamiltonian} by:
\begin{equation}
  \tensorial{J}_{i,j}^q(S) = \frac{\tensorial{J}_{i,j}^c}{S^2}, \quad d_{2}^q(S) = \frac{d_{2}^c}{S^2}, \quad d_{4}^q(S) = \frac{d_{4}^c}{S^4};
\end{equation}
(cf.\ Ref.~\cite{QuantumHematite}).
The dispersion of the fully relativistic Hamiltonian can be obtained by numerical bosonic diagonalisation of \eq~\eqref{eq:HP_matrix}.
In the isotropic limit, given by \eq~\eqref{eq:HP_ana}, an analytical solution can be found as well.

The Holstein--Primakoff transformation yields a $2N \times 2N$ spin wave matrix for every $k$-point, where $N$ is the number of magnetic sublattices. For hematite this gives:
\begin{widetext}
{\tiny
  \begin{align}
    M(\vec{k}) =
    \begin{pmatrix}
      A_{A,A}(\vec{k})+C_A & B_{A,A}(\vec{k}) & A_{A,B}(\vec{k}) & B_{A,B}(\vec{k}) & A_{A,C}(\vec{k}) & B_{A,C}(\vec{k}) & A_{A,D}(\vec{k}) & B_{A,D}(\vec{k}) \\
      -B_{A,A}^\star (-\vec{k}) & -A^\star _{A,A}(-\vec{k})-C^\star _A & -B_{A,B}^\star (-\vec{k}) & -A_{A,B}^\star (-\vec{k}) & -B_{A,C}^\star (-\vec{k}) & -A_{A,C}^\star (-\vec{k}) & -B_{A,D}^\star (-\vec{k}) & -A_{A,D}^\star (-\vec{k}) \\
      A_{B,A}(\vec{k}) & B_{B,A}(\vec{k}) & A_{B,B}(\vec{k})+C_B & B_{B,B}(\vec{k}) & A_{B,C}(\vec{k}) & B_{B,C}(\vec{k}) & A_{B,D}(\vec{k}) & B_{B,D}(\vec{k}) \\
      -B_{B,A}^\star (-\vec{k}) & -A^\star _{B,A}(-\vec{k}) & -B_{B,B}^\star (-\vec{k}) & -A_{B,B}^\star (-\vec{k})-C^\star _B & -B_{B,C}^\star (-\vec{k}) & -A_{B,C}^\star (-\vec{k}) & -B_{B,D}^\star (-\vec{k}) & -A_{B,D}^\star (-\vec{k}) \\
      A_{C,A}(\vec{k}) & B_{C,A}(\vec{k}) & A_{C,B}(\vec{k}) & B_{C,B}(\vec{k}) & A_{C,C}(\vec{k})+C_C & B_{C,C}(\vec{k}) & A_{C,D}(\vec{k}) & B_{C,D}(\vec{k}) \\
      -B_{C,A}^\star (-\vec{k}) & -A^\star _{C,A}(-\vec{k}) & -B_{C,B}^\star (-\vec{k}) & -A_{C,B}^\star (-\vec{k}) & -B_{C,C}^\star (-\vec{k}) & -A_{C,C}^\star (-\vec{k})-C^\star _C & -B_{C,D}^\star (-\vec{k}) & -A_{C,D}^\star (-\vec{k}) \\
      A_{D,A}(\vec{k}) & B_{D,A}(\vec{k}) & A_{D,B}(\vec{k}) & B_{D,B}(\vec{k}) & A_{D,C}(\vec{k}) & B_{D,C}(\vec{k}) & A_{D,D}(\vec{k})+C_C & B_{D,D}(\vec{k}) \\
      -B_{D,A}^\star (-\vec{k}) & -A^\star _{D,A}(-\vec{k}) & -B_{D,B}^\star (-\vec{k}) & -A_{D,B}^\star (-\vec{k}) & -B_{D,C}^\star (-\vec{k}) & -A_{D,C}^\star (-\vec{k}) & -B_{D,D}^\star (-\vec{k}) & -A_{D,D}^\star (-\vec{k})-C^\star _D \\
    \end{pmatrix}  , \label{eq:HP_matrix}
  \end{align}
}\end{widetext}
where $A_{X,Y}, B_{X,Y}$ and $ C_{X}$ are the following functions for the collinear phase of hematite
{\tiny
\begin{align}
  A_{Z,X}(\vec{k}) &=  \frac{S}{4}
  \left( \sum_l  \exp\left( \iu \vec{k} \left( \vec{r}_{0X} - \vec{r}_{lZ}\right)\right)  \left(-  \vec{u}^\star   \Tilde{{\tensorial{J}}}_{0X,lZ}  \vec{u} -     \vec{u}  \Tilde{{\tensorial{J}}}_{lZ,0X}   \vec{u}^\star  \right) \right),  \label{eq:funcA}\\
  B_{Z,X}(\vec{k}) &=  \frac{S}{4}
  \left( \sum_l  \exp\left( iu \vec{k} \left( \vec{r}_{0X} - \vec{r}_{lZ}\right)\right)  \left( -     \vec{u} \Tilde{{\tensorial{J}}}_{lZ,0X}   \vec{u}  -    \vec{u} \Tilde{{\tensorial{J}}}_{0X,lZ}   \vec{u} \right) \right),  \label{eq:funcB}\\
  C_{Z} &=  \frac{S}{2} \sum_{nX}\left(  \vec{v}    \Tilde{{\tensorial{J}}}      _{0Z,nX}  \vec{v} +   \vec{v}    \Tilde{{\tensorial{J}}}_{nX,0Z}       \vec{v}\right)  + S \cdot 2 \cdot     \Tilde{{\tensorial{K}}}_{Z}^{zz}       + S^3  \cdot 4 \cdot    \Tilde{{\tensorial{L}}}_{Z}^{z,z,z,z}       .   \label{eq:funcC}
\end{align}
}The subscripts denote the magnetic sublattices $A,B,C$ and $D$ and $S$ is the spin quantum number, which is obtained from the ab initio calculated magnetic moment $\mu = g \mu_\mathrm{B} S$. The Holstein--Primakoff transformation requires a ferromagnetic groundstate, which is obtained by rotating the sublattices $B$ and $C$ and their respective interactions, in those formulas this is marked by the tilde above the tensors ( The sublattices $A$ and $D$ are only "rotated" by the identity matrix). The sum over the interactions $l$ scales each interaction by the respective distance, for which the distance vectors $\vec{r}_{0X}-\vec{r}_{lZ}$ are needed and also its projection on the $k$-direction. Defined are also the vectors $\vec{u} = (1,\iu,0)$ and  $\vec{v} = (0,0,1)$~\cite{TothLake2015_SpinW}. The second and fourth order on-site anisotropies are given in their tensorial form ${\tensorial{K}}$ and ${\tensorial{L}}$.

The altermagnetic splitting is a non-relativistic effect, therefore one can simplify this $8\times 8$ matrix, by removing every non-isotropic exchange. We keep the second- and fourth-order on-site anisotropies to guarantee the correct magnetic groundstate. In this limit the $8\times 8$ matrix can be separated into two $4\times 4$ matrices via a similarity transformation. This ensures that the matrices have the same characteristic polynomial, hence the same eigenvalues
\begin{align}
    M_{\text{iso.}}(\vec{k}) =  \left(\begin{array}{@{}cc|cc@{}}
    \pll & \pll & \pll & \pll \\
    & \twotofour{$\xi(\vec{k})$} & & \twotofour{$0_{4\times4}$} \\
    \hline
    & & & \\
    & \twotofour{$0_{4\times4}$} & &  \twotofour{$-\xi(\vec{k})$}
  \end{array}\right) .
\end{align}
Since this is a block-diagonal matrix every eigenvalue of the $\xi(\vec{k})$ matrix will also show with the opposite sign in the $-\xi(\vec{k})$ matrix block. In this limit it is therefore sufficient to only study the $4\times4$ matrix $\xi(\vec{k})$
{\tiny
\begin{align*}
  \xi(\vec{k}) = \begin{pmatrix}
    A_{A,A}(\vec{k})+C_A &  B_{A,B}(\vec{k}) &  B_{A,C}(\vec{k}) & A_{A,D}(\vec{k})           \\
    -B_{A,B}^\star (\vec{k}) & -A_{A,A}(\vec{k}) -C_A  & -A_{B,C}(\vec{k}) & -B_{A,C}(\vec{k})   \\
    -B_{A,C}^\star (\vec{k}) & -A_{B,C}^\star (\vec{k})&  -A_{A,A}(\vec{k}) -C_A &  -B_{A,B}(\vec{k})   \\
    A_{A,D}^\star (\vec{k}) & B_{A,C}^\star (\vec{k})  & B_{A,B}^\star (\vec{k})  & A_{A,A}(\vec{k})+C_A  \\
  \end{pmatrix}.
\end{align*}
}From this it is straightforward to establish the characteristic polynomial
\begin{align}
  p(\lambda) = \lambda^4 + c_1(\vec{k}) \lambda^3 + c_2(\vec{k}) \lambda^2 + c_3(\vec{k}) \lambda + c_4(\vec{k}). \label{eq:charpoly}
\end{align}
Obviously $c_1(\vec{k}) = 0 \ \forall\ \vec{k}$, which makes this a depressed quartic. This can be solved for the trivial case of $c_3 = 0$ by
\begin{align}
  \lambda_{\pm, \pm}(\vec{k}) = \pm \sqrt{\frac{-c_2(\vec{k}) \pm \sqrt{c_2(\vec{k})^2 - 4 \cdot c_4(\vec{k})}}{2}},
\end{align}
and in the case of $c_3 \neq 0$ by the Ferrari method
\begin{align}
  \lambda_{\pm,\pm} = \frac{1}{2}\Biggl( &\pm \sqrt{2q(\vec{k})-c_2(\vec{k})}  \nonumber \\
   &\pm  \sqrt{ \frac{2c_3(\vec{k})}{\sqrt{2q(\vec{k})-c_2(\vec{k})}} -2q(\vec{k}) -c_2(\vec{k}) } \Biggr), \label{eq:HP_ana}
\end{align}
but in order to solve this one needs to find $q$, which is defined as the solution to:
\begin{align}
  q(\vec{k}) &= \frac{1}{3} \left( \frac{c_2(\vec{k})}{2} + 3w - \frac{p(\vec{k})}{w} \right),\\
  w(\vec{k}) &= \sqrt[3\; ]{-\frac{g(\vec{k})}{2}+\sqrt{\frac{g^2(\vec{k})}{4}+\frac{p^3(\vec{k})}{27}}},
\end{align}
with
\begin{align}
  p(\vec{k}) &= - \frac{c_2^2(\vec{k})}{12} - c_4(\vec{k}), \\
  g(\vec{k}) &= - \frac{c_2^3(\vec{k})}{108} + \frac{c_2(\vec{k})c_4(\vec{k})}{3} - \frac{c_3^2(\vec{k})}{8}.
\end{align}
This gives us an analytical solution for the magnon dispersion of hematite in the collinear phase.
With the definitions from \eqref{eq:funcA}--\eqref{eq:funcC}, and the additional shorthands
\begin{equation}
  a =  A_{A,A},\ b =  B_{A,B},\ c =  B_{A,C},\ d =  A_{A,D},\ e =  -A_{B,C}
  ,
\end{equation}
the $c_3$ coefficient can then be expressed as
\begin{equation}
  c_3 = \frac{2}{\iu \hbar}  \left( b^\star  \left(c^\star  d   +  e^\star c \right)- \text{c.c.} - a \left( e^\star e  - d^\star d\right)     \right)
  .
\end{equation}

\end{document}